\documentclass[aps,prb,twocolumn,notitlepage,floatfix,amsmath,amssymb,superscriptaddress]{revtex4-2}

\usepackage[T1]{fontenc}
\usepackage{bm}
\usepackage{graphicx}
\usepackage{booktabs}
\usepackage{mathtools}
\usepackage{microtype}
\usepackage[colorlinks=true,linkcolor=blue,citecolor=blue,urlcolor=blue]{hyperref}
\graphicspath{{figures/}}

\newcommand{\ii}{\mathrm{i}}
\newcommand{\dd}{\mathrm{d}}
\newcommand{\mcL}{\mathcal{L}}
\newcommand{\mcC}{\mathcal{C}}
\newcommand{\mcD}{\mathcal{D}}

\newcommand{\ellthree}{\ell_{3}}
\newcommand{\ve}{\nu_{\mathrm e}}
\newcommand{\vH}{\nu_{\mathrm H}}
\newcommand{\ke}{\kappa_{\mathrm e}}
\newcommand{\kH}{\kappa_{\mathrm H}}
\newcommand{\order}{\mathcal{O}}

\begin{document}

\title{Fourth-order closure obstruction and chiral nonlocality in circular kinetic magnetotransport}

\author{P. Shubham Parashar}
\email{psparash@ucsd.edu}
\affiliation{Department of Physics, University of California, San Diego, La Jolla, California 92093, USA}

\date{July 2026}

\begin{abstract}
Closing an angular moment hierarchy at the stress level omits a definite back-action from higher Fermi-surface harmonics.  For a circular two-dimensional Fermi surface, streaming changes angular momentum by one, so the shortest omitted sequence, $1\!\to\!2\!\to\!3\!\to\!2\!\to\!1$, adds a fourth-order term to the current eigenvalue,
$\Lambda(q)=\gamma_1+\nu q^2-\kappa_4q^4+\cdots$, with $\nu=v_F^2/(4\gamma_2)$ and $\kappa_4=\nu^2/\gamma_3$.  We call this missing operator term the fourth-order closure obstruction.  Its gradient expansion is controlled when $\nu q^2/\gamma_3\ll1$.  Circular symmetry carries the same coefficient into a radial bi-Laplacian within each conserved angular-momentum block, and retaining $m=3$ exactly, without a gradient expansion, amplifies higher radial modes monotonically.  At zero field, positive collision rates exclude real-wave-number poles and response zeros; an equal-rate tail gives a square-root completion.  A magnetic field makes the coefficient chiral, produces a Hall sign reversal, and enhances it when the $m=3$ harmonic is long lived.  In the collisionless high-field limit, the complete hierarchy becomes a Bessel pole--zero ladder, while finite closures form rational approximants to it.  The result separates a controlled low-gradient coefficient from its geometry- and field-dependent finite-wave-number completion.
\end{abstract}

\maketitle

\section{Introduction and conceptual overview}
\label{sec:introduction}

A moment closure replaces an infinite kinetic hierarchy by equations for a selected set of moments.  Exact elimination of the discarded moments does not make their dynamics disappear: their response returns as an operator acting on the retained sector.  We use \emph{closure obstruction} for the leading returned operator that prevents a chosen truncation from reproducing the exact retained-sector dynamics.  Chapman--Enskog, Grad, projection-operator, and systematic moment constructions express such feedback as higher gradients, memory, or nonlocal constitutive terms~\cite{ChapmanCowling,Burnett1936,Grad1949,Feshbach1958,Zwanzig1961,Mori1965,Levermore1996}.  Burnett-order corrections are therefore classical, while unrestricted finite-order polynomial equations can become unstable outside their long-wavelength domain~\cite{Bobylev1982,Uribe2000}.  Regularized moment equations provide one established way to preserve higher-moment information without extrapolating the bare polynomial to all wavelengths~\cite{StruchtrupTorrilhon2003}.

Here the question is concrete: what is the leading operator term omitted when a two-dimensional angular hierarchy is closed at the stress level?  The harmonics $m=0,\pm1,\pm2$ encode density, current, and stress, while $|m|\ge3$ describe finer distortions of the distribution~\cite{PhysicalKinetics,LucasFong2018,Ledwith2019,HofmannGran2023,NilssonGranHofmann2025}.  This density--current--stress sector is the kinetic content of the Navier--Stokes descriptions used for nonlocal viscous flow and electron backflow~\cite{Torre2015,LevitovFalkovich2016,Bandurin2016,LucasFong2018}.  Streaming couples neighboring harmonics.  The shortest sequence that leaves the retained current sector, visits the lowest discarded harmonic, and returns is
\begin{equation}
 1\longrightarrow2\longrightarrow3\longrightarrow2\longrightarrow1.
 \label{eq:introPath}
\end{equation}
For a plane wave, each streaming vertex carries a factor $\ii v_Fq/2$.  Four vertices therefore produce a fourth-order contribution.  Equivalently, the exact elimination may be viewed as a virtual excursion through $m=3$: current drives stress, stress excites the discarded distortion, and that distortion feeds back on the current.

Closing at stress level consequently changes the current operator first at fourth derivative order.  For a plane wave, the current eigenvalue is
\begin{align}
 \Lambda(q)
 &=\gamma_1+\nu q^2-\kappa_4q^4+\order(q^6),
 \nonumber\\
 \nu&=\frac{v_F^2}{4\gamma_2},
 \qquad
 \kappa_4=\frac{\nu^2}{\gamma_3},
 \nonumber\\
 \Xi(q)&\equiv\frac{\nu q^2}{\gamma_3}\ll1.
 \label{eq:introFourier}
\end{align}
Here $\nu$ is the kinematic viscosity, $\kappa_4$ is fixed by the relaxation rate of the lowest discarded harmonic rather than introduced phenomenologically, and $\Xi$ is the ratio of the fourth-order correction to the viscous term.

Equation~\eqref{eq:introFourier} is written for Cartesian plane waves.  A bare $q^4$ label is therefore coordinate specific and does not by itself establish what survives in polar coordinates.  Section~\ref{subsec:principalSymbol} recasts the fourth-derivative term through its coordinate-independent highest-derivative part, or principal symbol, after the transport meaning of $\nu$, $\kappa_4$, and $\Xi$ has been established.  In circular geometry that invariant term becomes a radial bi-Laplacian within the same conserved total-angular-momentum block.

The same fourth-order term also has a response-level meaning.  Through order $q^2$, a change of viscosity and a change of the second spatial moment of a smooth source produce the same broadening of the source-to-current map.  Once the source family is calibrated, the fourth-order coefficient is the first term that distinguishes the closed current response from the exact one.  The statement concerns the response map, not a claim that the fourth-order term must dominate numerically in every hydrodynamic regime.

Circular symmetry makes the coordinate-invariant statement concrete.  Total angular momentum $J=n+m$ is conserved, and exact radial ladder identities return the $m=3$ feedback to the same $J$ block as the bi-Laplacian $-\kappa_4\Delta_{J-1}^2$.  The closure error therefore does not first announce itself through a new angular channel; it reweights radial multipoles within the channel already driven by the source.

A magnetic field gives the discarded harmonic a phase as well as a lifetime, so the same operator becomes chiral.  Hall viscosity is an established transport coefficient of mesoscopic electron flow~\cite{Scaffidi2017,Pellegrino2017,Berdyugin2019}.  The coefficient derived here is not an additional independent viscosity; it is the leading kinetic correction to the Hall-viscous current operator.  In the dc helicity sectors,
\begin{align}
 \nu_{\pm}&=\frac{v_F^2}{4\bar\lambda_{\pm2}},
 \nonumber\\
 \kappa_{\pm}&=\frac{v_F^4}
 {16\bar\lambda_{\pm2}^{2}\bar\lambda_{\pm3}},
 \qquad
 \bar\lambda_{\pm m}=\gamma_m\pm\ii m\omega_c.
 \label{eq:introChiral}
\end{align}
The Hall component of $\kappa_\pm$ changes sign with field.  When the $m=3$ mode is anomalously long lived, its low-field branch is enhanced by $1/\gamma_3$ and peaks near $3\omega_c=\gamma_3$.  This connects the closure obstruction to odd-harmonic physics in two-dimensional Fermi liquids and to circular magnetotransport~\cite{Ledwith2019,HofmannGran2023,NilssonGranHofmann2025,Moiseenko2025,Rostami2025,ParasharFogler2026}.

The controlled coefficient is not a finite-wave-number theory.  At zero field, positive collision rates exclude real-$q$ poles and response zeros.  In collisionless high field, magnetic precession instead reorganizes the complete hierarchy into
\begin{equation}
 G_\infty(k)
 =\frac{2J_1(kR_c)}{\ii\omega_c kR_c J_0(kR_c)}.
 \label{eq:introBessel}
\end{equation}
The zeros of the denominator are current poles, while zeros of the numerator are response zeros.  Finite closures are rational approximants to this Bessel pole--zero ladder.  The $M=3$ values $kR_c=\sqrt6$ and $\sqrt{24}$ are the lowest approximants to the first pole and response zero, not universal predictions of the fourth-order gradient theory.

A previous source--operator analysis treated a Hall-odd $q^4$ closure correction as an undetermined systematic in scanning-probe inference~\cite{ParasharScanning2026}.  The present work derives that correction as a property of the kinetic operator, without assuming a probe or inverse problem.  The following sections derive the Schur complement, establish its circular and chiral forms, and separate the controlled low-gradient coefficient from the finite-$k$ kinetic completion.

\section{Angular hierarchy and the shortest omitted feedback path}
\label{sec:hierarchy}

\subsection{Conventions and harmonic chain}

We consider a linearized kinetic equation on a circular two-dimensional Fermi surface,
\begin{equation}
 \left[
 \partial_t+v_F\hat{\bm p}(\vartheta)\!\cdot\!\nabla
 +\omega_c\partial_\vartheta+\mcC
 \right]\phi(\bm r,\vartheta,t)=S(\bm r,\vartheta,t),
 \label{eq:boltzmann}
\end{equation}
where $\hat{\bm p}=(\cos\vartheta,\sin\vartheta)$.  We define the signed cyclotron frequency by $\dot\vartheta=\omega_c$, so positive $\omega_c$ advances the momentum angle counterclockwise.  For carrier charge $q_{\rm car}$ in a field $B_z\hat{\bm z}$,
\begin{equation}
 \omega_c=-\frac{q_{\rm car}B_zv_F}{p_F},
 \label{eq:cyclotronConvention}
\end{equation}
which reduces to $-q_{\rm car}B_z/m^\ast$ for a parabolic band.  All helicity and Hall signs below refer to this convention.

Rotational invariance makes the collision operator block diagonal in angular momentum.  We additionally project each angular block onto one radial or energy eigenmode, represented by the scalar rate
\begin{equation}
 \mcC e^{\ii m\vartheta}=\gamma_m e^{\ii m\vartheta}.
 \label{eq:collisionSpectrum}
\end{equation}
The rates may include momentum relaxation and momentum-conserving collisions, and no equality among them is assumed.  Among the discarded rates, the fourth-order coefficient depends only on $\gamma_3$; rates $\gamma_{m\ge4}$ first enter at sixth derivative order.

For a plane wave $e^{\ii\bm q\cdot\bm r-\ii\Omega t}$, choose $\bm q=q\hat{\bm x}$ and expand $\phi=\sum_m f_m e^{\ii m\vartheta}$.  The hierarchy is
\begin{align}
 \lambda_m f_m+\ii t_q(f_{m-1}+f_{m+1})&=S_m,
 \label{eq:hierarchy}\\
 \lambda_m&=\gamma_m-\ii\Omega+\ii m\omega_c,
 \label{eq:lambda}\\
 t_q&=\frac{v_Fq}{2}.
\end{align}
The retarded prescription is implicit.  All explicit coefficients, Green functions, and resummations below are evaluated in the dc limit $\Omega=0$; the frequency is retained in Eqs.~\eqref{eq:hierarchy} and \eqref{eq:lambda} only to define the retarded operator.  At $B=0$, the $m$ and $-m$ sectors are conjugate copies.  At finite field it is convenient to use helicity chains with $\lambda_{\pm m}=\gamma_m-\ii\Omega\pm\ii m\omega_c$.

\subsection{Stress closure and the unexpanded $M=3$ current operator}

Let $P_2$ project onto $|m|\le2$ and $Q=1-P_2$.  Exact elimination of the discarded sector gives the Schur complement
\begin{equation}
 \mcL_{\mathrm{eff}}
 =P_2\mcL P_2
 -P_2\mcL Q\,(Q\mcL Q)^{-1}Q\mcL P_2.
 \label{eq:mainSchur}
\end{equation}
The second term is the returned operator described in the Introduction.  For a source in the current sector, the $M=2$ stress closure gives
\begin{equation}
 \Lambda_2(q)=\lambda_1+\frac{t_q^2}{\lambda_2}.
 \label{eq:Lambda2general}
\end{equation}
Retaining the lowest discarded harmonic without a gradient expansion gives the continued fraction
\begin{equation}
 \boxed{
 \Lambda_3(q)
 =\lambda_1+
 \frac{t_q^2}{\lambda_2+t_q^2/\lambda_3}.
 }
 \label{eq:Lambda3general}
\end{equation}
Here and below, ``unexpanded $M=3$'' means exact within the $|m|\le3$ truncation, with no gradient expansion; it does not mean exact for the infinite hierarchy.
At zero external frequency and zero field,
\begin{equation}
 \Lambda_2(q)=\gamma_1+\nu q^2,
 \qquad
 \nu=\frac{v_F^2}{4\gamma_2},
 \label{eq:viscosity}
\end{equation}
while Eq.~\eqref{eq:Lambda3general} expands as
\begin{equation}
 \boxed{
 \Lambda_3(q)
 =\gamma_1+\nu q^2-\kappa_4q^4+\order(q^6),
 \qquad
 \kappa_4=\frac{\nu^2}{\gamma_3}.
 }
 \label{eq:q4expansion}
\end{equation}
The ratio of the fourth-order correction to the viscous term defines the expansion parameter
\begin{equation}
 \boxed{
 \Xi(q)\equiv
 \left|\frac{\kappa_4q^4}{\nu q^2}\right|
 =\frac{\nu q^2}{\gamma_3}
 =\frac{v_F^2q^2}{4\gamma_2\gamma_3}.
 }
 \label{eq:XiControl}
\end{equation}
The gradient expansion is controlled when $\Xi(q)\ll1$; for a circular radial mode one replaces $q^2$ by $k_{n\ell}^2$.  The relative sign is fixed by anti-Hermitian streaming.  The two extra streaming vertices carry the factor $(\ii q)^2=-q^2$, so the first correction opposes the $q^2$ viscous term.  Physically, the stress does not relax purely locally: it can propagate through the $m=3$ distortion before feeding back into the current.  Equation~\eqref{eq:q4expansion} is the low-$q$ expansion of the positive $M=3$ rational approximant
\begin{equation}
 \Lambda_3(q)=\gamma_1+
 \frac{\nu q^2}{1+\nu q^2/\gamma_3}.
 \label{eq:Lambda3B0}
\end{equation}
This approximant remains positive for real $q$ and avoids the spurious turnover of a standalone $q^4$ polynomial.  Its saturation at $\gamma_1+\gamma_3$, however, is a truncation artifact rather than a physical large-$q$ limit.  The equal-rate full-hierarchy completion in Eq.~\eqref{eq:B0exactTail} instead has $X_\infty(q)\sim v_F|q|/2$, recovering ballistic free-streaming growth.  Higher harmonics cannot change the coefficient in Eq.~\eqref{eq:q4expansion}: reaching $m=4$ and returning to the current requires six streaming vertices, so $\kappa_4$ is already the exact full-hierarchy coefficient at fourth order even though Eq.~\eqref{eq:Lambda3B0} is not the full finite-$q$ response.

\subsection{Zero-field positivity and an equal-rate completion}
\label{subsec:B0resummation}

At $B=\Omega=0$, let $z=t_q^2\ge0$.  For any finite truncation with positive relaxation rates, the unresolved tail is a Stieltjes continued fraction, that is, a continued fraction built from positive elements,
\begin{equation}
 X_2^{(M)}(z)=
 \cfrac{z}{\gamma_2+
 \cfrac{z}{\gamma_3+\cdots+
 \cfrac{z}{\gamma_M}}}.
 \label{eq:B0positiveFraction}
\end{equation}
Every denominator is positive, and therefore
\begin{equation}
 0\le X_2^{(M)}(z)\le\frac{z}{\gamma_2},
 \qquad
 \gamma_1\le\Lambda_M(q)\le
 \gamma_1+\frac{z}{\gamma_2}.
 \label{eq:B0positivityBounds}
\end{equation}
Any convergent semi-infinite positive-rate hierarchy inherits these bounds.  Hence the static current response $G(q)=1/\Lambda(q)$ has neither a real-wave-number pole nor a real-wave-number response zero at zero field.  This real-axis positivity is general; the singularity structure away from the real axis depends on the full rate spectrum.

A useful solvable benchmark is the special unresolved tail $\gamma_m=\gamma$ for every $m\ge2$.  This choice has $\gamma_2=\gamma_3$ and therefore does not describe odd-mode protection; its purpose is to expose one exact finite-$q$ completion.  Translation invariance along the tail makes its self-energy $X_\infty$ satisfy
\begin{equation}
 X_\infty(q)=\frac{t_q^2}{\gamma+X_\infty(q)}.
 \label{eq:B0fixedPoint}
\end{equation}
The branch selected by the finite positive convergents and analytic at $q=0$ is
\begin{equation}
 \boxed{
 X_\infty(q)=\frac12\left[\sqrt{\gamma^2+v_F^2q^2}-\gamma\right].
 }
 \label{eq:B0exactTail}
\end{equation}
The corresponding current eigenvalue is $\Lambda_\infty(q)=\gamma_1+X_\infty(q)$.  Its expansion is
\begin{equation}
 X_\infty(q)
 =\frac{v_F^2}{4\gamma}q^2
 -\frac{v_F^4}{16\gamma^3}q^4+\order(q^6)
 =\nu q^2-\frac{\nu^2}{\gamma}q^4+\order(q^6),
 \label{eq:B0exactExpansion}
\end{equation}
which reproduces Eq.~\eqref{eq:q4expansion} with $\gamma_2=\gamma_3=\gamma$.  At large wave number,
\begin{equation}
 X_\infty(q)=\frac{v_F|q|}{2}-\frac{\gamma}{2}+\order(|q|^{-1}),
 \label{eq:B0ballisticAsymptote}
\end{equation}
so the complete equal-rate hierarchy restores the unbounded free-streaming growth that any fixed moment truncation misses.  This benchmark has square-root branch points at $q=\pm\ii\gamma/v_F$.  Unequal rates can produce a more complicated off-axis analytic structure, but Eq.~\eqref{eq:B0positivityBounds} still excludes a real-axis pole--zero ladder.  The Bessel ladder found in Sec.~\ref{sec:bessel} is therefore a magnetic real-wave-number structure, not a generic consequence of retaining more moments.  Supplemental Material, Sec.~\ref{sec:suppB0Tail}, gives the continued-fraction derivation.

\subsection{Source width and viscosity enter the same $q^2$ response}
\label{subsec:sourceequivalence}

The fourth-order closure obstruction is most transparent in a source-to-current response.  Let an inversion-symmetric smooth source have the expansion
\begin{equation}
 S(q)=S_0\left[1-s_2q^2+s_4q^4+\order(q^6)\right].
 \label{eq:sourceExpansion}
\end{equation}
Writing $\alpha=\nu/\gamma_1$, Eqs.~\eqref{eq:q4expansion} and \eqref{eq:sourceExpansion} give
\begin{align}
 \frac{j(q)}{S_0/\gamma_1}
 &=1-(s_2+\alpha)q^2
 \nonumber\\
 &\quad+
 \left[
 s_4+\alpha s_2+\alpha^2+\frac{\kappa_4}{\gamma_1}
 \right]q^4+\order(q^6).
 \label{eq:responseJet}
\end{align}
At order $q^2$, the response identifies only the combination
\begin{equation}
 s_2+\frac{\nu}{\gamma_1}.
 \label{eq:q2combination}
\end{equation}
A viscosity change can therefore be absorbed by a change in the source second moment.  This is a statement about the source-to-current map, not about the homogeneous operator alone, and it concerns the low-$q$ Taylor expansion of the source-to-current map rather than a particular coordinate representation.

The fourth-order statement requires one qualification.  If the source is completely unknown, its fourth moment $s_4$ can also absorb a fourth-order operator change.  The closure coefficient becomes identifiable when the source family is calibrated or constrained so that $s_4$ is fixed once the lower source parameters are specified.  Once the allowed source-shape variations are fixed or projected out, $\kappa_4/\gamma_1$ is the first operator-specific response coefficient.  The Gaussian one-scale source used in the original Fourier calculation is one exact realization of this condition; the circular Green-function calculation below uses a declared smoothed-ring profile instead.

This statement identifies the first distinguishable operator direction, not its guaranteed numerical dominance.  Relative to the purely hydrodynamic $\alpha^2$ contribution in Eq.~\eqref{eq:responseJet}, the closure-specific term obeys
\begin{equation}
 \frac{\kappa_4/\gamma_1}{\alpha^2}=\frac{\gamma_1}{\gamma_3}.
 \label{eq:closureSensitivity}
\end{equation}
In the conventional hydrodynamic hierarchy $\gamma_1\ll\gamma_3$, it is parametrically suppressed; resolving it requires either an unusually slow $m=3$ mode or access to sufficient fourth-order spectral structure.

\subsection{Coordinate-independent form of the fourth-order term}
\label{subsec:principalSymbol}

The Fourier variable $q$ is tied to a Cartesian plane-wave representation.  To carry the result to another coordinate system, we use the principal symbol of the differential operator.  For an operator $A$, $\sigma_4(A)(\bm x,\bm\xi)$ denotes the component of its local symbol that is homogeneous of degree four in the local wavevector $\bm\xi$, obtained by replacing each derivative $\partial_j$ by $\ii\xi_j$ and retaining the fourth-degree part.  The position variable is $\bm x$.

The Schur complement in Eq.~\eqref{eq:mainSchur} is coordinate independent.  For constant isotropic rates, the four streaming vertices in Eq.~\eqref{eq:introPath} give the leading omitted current-sector operator
\begin{equation}
 \delta\mcL_j^{(4)}=-\kappa_4\Delta^2
 \label{eq:deltaLflat}
\end{equation}
on each scalar helicity component of the current in flat space.  Its fourth-order principal symbol is
\begin{equation}
 \boxed{
 \sigma_4(\delta\mcL_j)(\bm x,\bm\xi)
 =-\kappa_4|\bm\xi|^4.
 }
 \label{eq:principalSymbol}
\end{equation}
This is the coordinate-invariant content of the Fourier $-\kappa_4q^4$ term.  Coordinate transformations can generate lower-order derivative and connection terms, but they do not alter this scalar fourth-order symbol.  Circular coordinates expose how the same coefficient is organized by rotational symmetry.

\section{Circular factorization and radial multipoles}
\label{sec:circular}

Rotational symmetry lets the same fourth-order coefficient be written in closed form in polar coordinates: total angular momentum is conserved, and each angular-momentum block can be treated separately.

\subsection{Conserved total angular momentum}

Write the spatial and Fermi-surface angles as $\varphi$ and $\vartheta$.  The distribution is expanded as
\begin{equation}
 \phi(r,\varphi,\vartheta)
 =\sum_{J,m}f_m^{(J)}(r)
 e^{\ii(J-m)\varphi}e^{\ii m\vartheta}.
 \label{eq:circularExpansion}
\end{equation}
Rotational symmetry conserves
\begin{equation}
 J=n+m,
 \label{eq:Jconservation}
\end{equation}
where $n$ is spatial angular momentum.  The polar streaming operator is
\begin{align}
 \hat{\bm p}\!\cdot\!\nabla
 &=\frac12e^{\ii(\vartheta-\varphi)}
 \left(\partial_r-\frac{\ii}{r}\partial_\varphi\right)
 \nonumber\\
 &\quad+
 \frac12e^{-\ii(\vartheta-\varphi)}
 \left(\partial_r+\frac{\ii}{r}\partial_\varphi\right).
 \label{eq:polarStreaming}
\end{align}
It changes $m$ by one while leaving $J$ fixed.  Define the radial ladders
\begin{equation}
 \mcD_m^+=\partial_r+\frac{J-m}{r},
 \qquad
 \mcD_m^-=\partial_r-\frac{J-m}{r}.
 \label{eq:ladders}
\end{equation}
Their first exact identity is
\begin{equation}
 \boxed{
 \mcD_{m+1}^-\mcD_m^+=\Delta_{J-m},
 \qquad
 \Delta_n=\partial_r^2+\frac1r\partial_r-\frac{n^2}{r^2}.
 }
 \label{eq:ladderIdentity}
\end{equation}

\subsection{Unexpanded $M=3$ Schur complement in circular geometry}

At $B=0$, eliminating $m=3$ inside one $J$ block gives the unexpanded stress denominator
\begin{equation}
 \gamma_2-\frac{v_F^2}{4\gamma_3}\Delta_{J-2}.
 \label{eq:stressDenominator}
\end{equation}
The current operator retained through $m=3$ is therefore
\begin{equation}
 \boxed{
 \mcL_{j,3}^{(J)}
 =\gamma_1-\frac{v_F^2}{4}
 \mcD_2^-
 \left(
 \gamma_2-\frac{v_F^2}{4\gamma_3}\Delta_{J-2}
 \right)^{-1}
 \mcD_1^+.
 }
 \label{eq:exactCircularSchur}
\end{equation}
Expanding the inverse gives the $M=2$ viscous term and the first correction.  The second exact ladder identity is
\begin{equation}
 \boxed{
 \mcD_2^-\Delta_{J-2}\mcD_1^+
 =\Delta_{J-1}^2.
 }
 \label{eq:bilaplacianIdentity}
\end{equation}
Hence
\begin{equation}
 \boxed{
 \mcL_j^{(J)}
 =\gamma_1-\nu\Delta_{J-1}
 -\kappa_4\Delta_{J-1}^2
 +\order(\nabla^6).
 }
 \label{eq:circularOperator}
\end{equation}

This factorization is the circular content of Eq.~\eqref{eq:principalSymbol}.  The closure error remains inside the same conserved $J$ block.  In particular, a uniform electric drive selects $J=\pm1$; the omitted $m=3$ mode feeds back into those blocks rather than announcing itself through a new total-angular-momentum channel.

For a rotationally invariant domain with radial eigenfunctions
\begin{equation}
 \Delta_nu_{n\ell}=-k_{n\ell}^2u_{n\ell},
 \label{eq:radialEigenfunctions}
\end{equation}
Eq.~\eqref{eq:exactCircularSchur} reduces mode by mode to Eq.~\eqref{eq:Lambda3B0}.  The hydrodynamic and first closure terms scale as $k_{n\ell}^2$ and $k_{n\ell}^4$, respectively.  The first independent closure information is therefore carried by the higher radial multipoles of the same angular response.

\section{Radial Green function of the $M=3$ Schur complement}
\label{sec:green}

\subsection{Local plus screened decomposition}

The controlled finite-$k$ object is the unexpanded $M=3$ response function, not the isolated fourth-order partial differential equation.  We refer to the static response functions below as resolvents.  Set $z=k^2$ and define
\begin{equation}
 R_2(z)=\frac1{\gamma_1+\nu z},
 \qquad
 R_3(z)=\frac1{\gamma_1+\nu z/(1+\nu z/\gamma_3)}.
 \label{eq:resolvents}
\end{equation}
The second expression admits the exact decomposition
\begin{equation}
 \boxed{
 R_3(z)=\frac1{\gamma_1+\gamma_3}
 +\frac{\gamma_3^2}{\nu(\gamma_1+\gamma_3)^2}
 \frac1{z+\mu_3^2},
 }
 \label{eq:partialFraction}
\end{equation}
where
\begin{equation}
 \mu_2^2=\frac{\gamma_1}{\nu},
 \qquad
 \mu_3^2=\frac{\gamma_1\gamma_3}{\nu(\gamma_1+\gamma_3)}.
 \label{eq:screeningMasses}
\end{equation}
The $M=2$ screening length is the Gurzhi length,
\begin{equation}
 \ell_G\equiv\mu_2^{-1}=\sqrt{\frac{\nu}{\gamma_1}},
 \label{eq:GurzhiLength}
\end{equation}
namely the momentum-relaxation-limited viscous screening length~\cite{Gurzhi1968,LucasFong2018}.
Indeed,
\begin{equation}
 \mu_2^2-\mu_3^2
 =\frac{\gamma_1^2}{\nu(\gamma_1+\gamma_3)}>0,
 \label{eq:massDifference}
\end{equation}
so the screening-length ratio is
\begin{equation}
 \frac{\mu_2}{\mu_3}
 =\sqrt{1+\frac{\gamma_1}{\gamma_3}}>1.
 \label{eq:screeningLengthRatio}
\end{equation}
The $M=3$ nonlocal component is therefore longer ranged than the $M=2$ viscous kernel, although the increase is only $\simeq\gamma_1/(2\gamma_3)$ in the conventional hierarchy $\gamma_1\ll\gamma_3$.  The strictly monotone spectral gain below, rather than a parametrically enlarged screening length, is the robust effect.

For a source $s_n(r)$ in spatial angular channel $n=J-1$, the radial Green functions, with measure $r'\dd r'$, are
\begin{equation}
 G_2^{(n)}(r,r')
 =\frac1\nu I_n(\mu_2r_<)K_n(\mu_2r_>),
 \label{eq:G2}
\end{equation}
\begin{align}
 G_3^{(n)}(r,r')
 &=\frac1{\gamma_1+\gamma_3}
 \frac{\delta(r-r')}{r}
 \nonumber\\
 &\quad+
 \frac{\gamma_3^2}{\nu(\gamma_1+\gamma_3)^2}
 I_n(\mu_3r_<)K_n(\mu_3r_>).
 \label{eq:G3}
\end{align}
The lowest discarded harmonic therefore generates two pieces: a local copy of the circular source and a screened Bessel response with the length ratio in Eq.~\eqref{eq:screeningLengthRatio}.

\subsection{Monotone transfer toward higher radial modes}

Define the gain of the unexpanded $M=3$ response relative to the $M=2$ closure,
\begin{equation}
 Q(z)=\frac{R_3(z)}{R_2(z)}
 =\frac{\Lambda_2(z)}{\Lambda_3(z)}.
 \label{eq:gain}
\end{equation}
Direct differentiation gives
\begin{equation}
 \boxed{
 \frac{\dd Q}{\dd z}
 =\frac{\nu^2z\,[2\gamma_1\gamma_3+
 \nu z(\gamma_1+\gamma_3)]}
 {[\gamma_1\gamma_3+\nu z(\gamma_1+\gamma_3)]^2}>0
 }
 \label{eq:gainDerivative}
\end{equation}
for $z>0$.  The unexpanded $M=3$ gain is therefore strictly increasing with radial wave number.  This is stronger than a source-specific sharpening effect: it orders the entire radial spectrum.  For any source that occupies more than one radial mode, normalizing the two responses produces a monotone spectral reweighting of the $M=2$ spectrum toward larger $k^2$.  Every increasing spectral moment, including $\langle k^2\rangle$ and $\langle k^4\rangle$, consequently increases.

\begin{figure}[t]
 \centering
 \includegraphics[width=0.96\columnwidth]{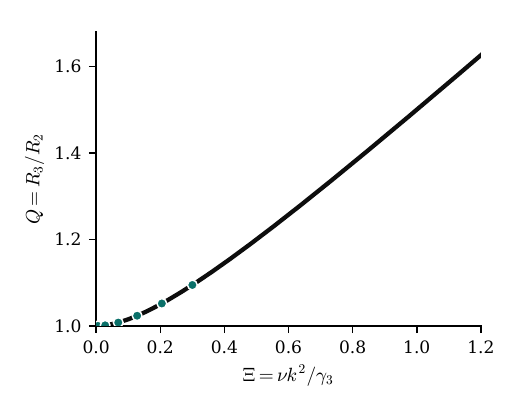}
 \caption{\label{fig:radial}\textbf{Radial-mode gain generated by the lowest discarded harmonic.}  The ratio $Q(k^2)=R_3(k^2)/R_2(k^2)$ between the unexpanded $M=3$ Schur-complement response function and the $M=2$ hydrodynamic response function is strictly increasing for $k^2>0$.  The solid curve is Eq.~\eqref{eq:gain}; the symbols mark the declared low-gradient sample used in the Supplemental Material illustration, $\Xi_\ell\le0.30$.}
\end{figure}

The monotonicity statement is exact and source independent.  Figure~\ref{fig:radial} displays the analytic gain together with the discrete modes used in the controlled illustration of Supplemental Material, Sec.~\ref{sec:suppNumerics}.  Those symbols satisfy $\Xi_\ell=\nu k_\ell^2/\gamma_3\le0.30$.  The corresponding smoothed-ring profile and normalized mode-weight redistribution are deferred to Supplemental Material, Fig.~\ref{fig:suppRing}, where they are presented only as an illustration of the theorem.

\section{Chiral fourth-order coefficient at finite field}
\label{sec:chiral}

\subsection{Helicity-resolved coefficients}

A magnetic field gives the lowest discarded harmonic a phase as well as a lifetime, so its feedback on the current becomes chiral.  For the two circular current helicities, Eq.~\eqref{eq:lambda} becomes
\begin{equation}
 \lambda_{\pm m}=\gamma_m-\ii\Omega\pm\ii m\omega_c.
 \label{eq:lambdaHelicity}
\end{equation}
All results in this section are dc.  We therefore set
\begin{equation}
 \lambda_{\pm m}\longrightarrow
 \bar\lambda_{\pm m}\equiv\gamma_m\pm\ii m\omega_c.
 \label{eq:lambdaHelicityDc}
\end{equation}
The unexpanded $M=3$ current eigenvalue is
\begin{equation}
 \Lambda_\pm^{(3)}(k)
 =\bar\lambda_{\pm1}+
 \frac{(v_F^2/4)k^2}
 {\bar\lambda_{\pm2}+(v_F^2/4)k^2/\bar\lambda_{\pm3}}.
 \label{eq:finiteBcontinuedFraction}
\end{equation}
Its low-$k$ expansion is
\begin{equation}
 \Lambda_\pm(k)
 =\bar\lambda_{\pm1}+\nu_\pm k^2-\kappa_\pm k^4+\order(k^6),
 \label{eq:helicityExpansion}
\end{equation}
with
\begin{equation}
 \boxed{
 \nu_\pm=\frac{v_F^2}{4\bar\lambda_{\pm2}},
 \qquad
 \kappa_\pm=\frac{v_F^4}
 {16\bar\lambda_{\pm2}^2\bar\lambda_{\pm3}}.
 }
 \label{eq:helicityCoefficients}
\end{equation}
The finite-field expansion is controlled by the complex ratio
\begin{equation}
 \Xi_\pm(k)\equiv
 \frac{v_F^2k^2}{4\bar\lambda_{\pm2}\bar\lambda_{\pm3}},
 \qquad |\Xi_\pm(k)|\ll1.
 \label{eq:XiHelicity}
\end{equation}
At $B=0$, Eq.~\eqref{eq:XiHelicity} reduces to Eq.~\eqref{eq:XiControl}.  The usual even and Hall viscosities are the real and imaginary parts of $\nu_+$~\cite{Avron1995,HoyosSon2012,Pellegrino2017,Alekseev2016}.  The same decomposition defines a fourth-order even coefficient and a fourth-order Hall coefficient,
\begin{equation}
 \nu_+=\ve-\ii\vH,
 \qquad
 \kappa_+=\ke-\ii\kH.
 \label{eq:partsDefinition}
\end{equation}

Set $a=\gamma_2$, $c=\gamma_3$, and $\omega=\omega_c$.  The second-order coefficients are
\begin{equation}
 \ve=\frac{v_F^2a}{4(a^2+4\omega^2)},
 \qquad
 \vH=\frac{v_F^2\omega}{2(a^2+4\omega^2)}.
 \label{eq:nuParts}
\end{equation}
With
\begin{equation}
 D_\kappa=(a^2+4\omega^2)^2(c^2+9\omega^2),
 \label{eq:Dkappa}
\end{equation}
we obtain
\begin{equation}
 \ke=
 \frac{v_F^4[c(a^2-4\omega^2)-12a\omega^2]}
 {16D_\kappa},
 \label{eq:kappaEven}
\end{equation}
\begin{equation}
 \boxed{
 \kH=
 \frac{v_F^4\omega[3a^2+4ac-12\omega^2]}
 {16D_\kappa}.
 }
 \label{eq:kappaHall}
\end{equation}

\subsection{Field-driven sign reversal}

The Hall component changes sign at
\begin{equation}
 \boxed{
 \omega_{c,0}^2
 =\frac{\gamma_2(3\gamma_2+4\gamma_3)}{12}.
 }
 \label{eq:signReversal}
\end{equation}
For positive field, $\kH>0$ below $\omega_{c,0}$ and $\kH<0$ above it.  Because the spectral operator contains
\begin{equation}
 \vH k^2-\kH k^4,
 \label{eq:HallSpectralTerms}
\end{equation}
the fourth-order Hall term opposes Hall viscosity on the low-field branch and reinforces it after the sign reversal.  The sign change is a coefficient-level result and does not rely on extrapolating the gradient expansion to a finite-$k$ zero.

Figure~\ref{fig:chiral}a shows the sign reversal for several $\gamma_3/\gamma_2$.  The curves are normalized by their positive-branch maxima to separate the field dependence from the odd-mode enhancement of the absolute magnitude.

\subsection{Odd-mode enhancement}

\begin{figure*}[t]
 \centering
 \includegraphics[width=0.92\textwidth]{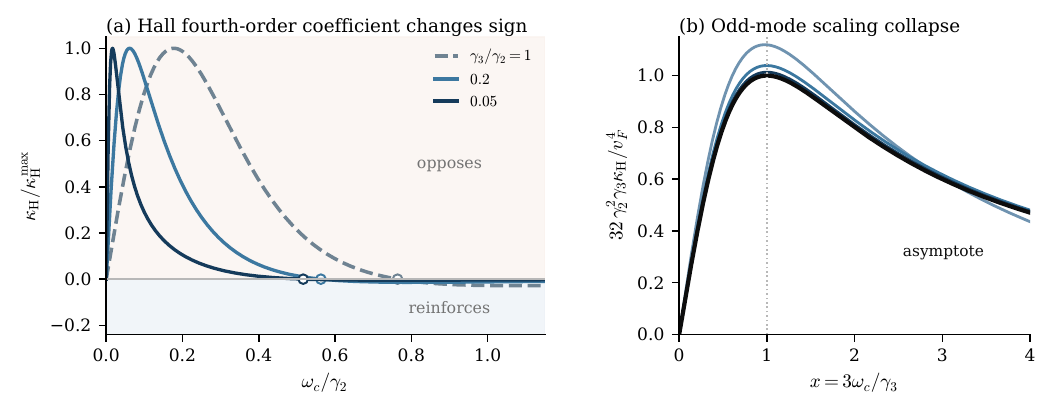}
 \caption{\label{fig:chiral}\textbf{Field dependence and odd-mode scaling of the chiral fourth-order closure coefficient.}  \textbf{(a)} Hall component $\kH$ for $\gamma_3/\gamma_2=1,0.2,$ and $0.05$, normalized by the maximum of its positive-field branch.  Open symbols mark the field at which $\kH$ changes sign.  In the dc helicity-resolved current operator $\Lambda_\pm(k)=\bar\lambda_{\pm1}+\nu_\pm k^2-\kappa_\pm k^4+\cdots$, positive $\kH$ makes the fourth-order Hall term oppose the Hall-viscous contribution, whereas negative $\kH$ makes the two contributions reinforce.  \textbf{(b)} Odd-mode scaling limit.  Curves for $\gamma_3/\gamma_2=0.1,0.03,$ and $0.01$ are plotted as $32\gamma_2^2\gamma_3\kH/v_F^4$ versus $x=3\omega_c/\gamma_3$.  They approach the universal function $2x/(1+x^2)$, shown in black, with its maximum at $x=1$.  The collapse displays the $1/\gamma_3$ enhancement when the $m=3$ harmonic is anomalously long lived.}
\end{figure*}

Two-dimensional Fermi liquids can possess anomalously long-lived odd angular harmonics~\cite{Ledwith2019,HofmannGran2023,NilssonGranHofmann2025}.  If $\gamma_3\ll\gamma_2$, the positive branch of Eq.~\eqref{eq:kappaHall} is concentrated near
\begin{equation}
 \boxed{
 \omega_c^{\max}=\frac{\gamma_3}{3}
 +\order\!\left(\frac{\gamma_3^3}{\gamma_2^2}\right).
 }
 \label{eq:peakField}
\end{equation}
At this field the $m=3$ precession rate $3\omega_c$ matches the relaxation rate $\gamma_3$.  The maximum is
\begin{equation}
 \boxed{
 \kH^{\max}
 =\frac{v_F^4}{32\gamma_2^2\gamma_3}
 \left[1+\frac{4\gamma_3}{3\gamma_2}+\order\!\left(\frac{\gamma_3^2}{\gamma_2^2}\right)\right].
 }
 \label{eq:kappaMaximum}
\end{equation}
Introducing
\begin{equation}
 x=\frac{3\omega_c}{\gamma_3},
 \label{eq:xOdd}
\end{equation}
the full positive branch has the asymptotic scaling form
\begin{equation}
 \boxed{
 \frac{32\gamma_2^2\gamma_3}{v_F^4}\kH
 \longrightarrow\frac{2x}{1+x^2}
 \qquad(\gamma_3/\gamma_2\to0).
 }
 \label{eq:oddScalingFunction}
\end{equation}
Figure~\ref{fig:chiral}b shows the collapse.

At the same field,
\begin{equation}
 \vH\simeq\frac{v_F^2\gamma_3}{6\gamma_2^2}.
 \label{eq:nuHallAtPeak}
\end{equation}
The ratio of the two low-$k$ Hall contributions is therefore
\begin{equation}
 \frac{\kH k^4}{\vH k^2}
 \simeq\frac{3}{16}(k\ellthree)^2,
 \qquad
 \ellthree=\frac{v_F}{\gamma_3}.
 \label{eq:ell3Crossover}
\end{equation}
Odd-mode protection moves the chiral nonlocal crossover to the long length $\ellthree$.  Formal equality occurs at $k\ellthree=4/\sqrt3$, where the gradient expansion is already becoming marginal.  The equation locates the scale at which $|\Xi_+|$ in Eq.~\eqref{eq:XiHelicity} becomes order unity; it should not be read as a controlled zero of a fourth-order partial differential equation.

At asymptotically high field,
\begin{equation}
 \vH\sim\frac{v_F^2}{8\omega_c},
 \qquad
 \kH\sim-\frac{v_F^4}{192\omega_c^3},
 \label{eq:highFieldAsymptotics}
\end{equation}
so
\begin{equation}
 \left|\frac{\kH k^4}{\vH k^2}\right|
 \sim\frac{(kR_c)^2}{24},
 \qquad
 R_c=\frac{v_F}{\omega_c}.
 \label{eq:kRcRatio}
\end{equation}
The cyclotron radius is the governing high-field length, but the equality of the two asymptotic magnitudes would occur at $kR_c=\sqrt{24}$, outside the controlled low-$k$ regime.  Section~\ref{sec:bessel} gives the finite-$k$ completion.

\section{Bessel resummation at finite cyclotron wave number}
\label{sec:bessel}

The fourth-order coefficient is a low-gradient statement.  To determine what replaces it at finite wave number, we return to the exact $M=3$ response before expanding in gradients.

\subsection{Why the $M=3$ denominator is not the finite-wave-number answer}

For a source in the positive current helicity, the unexpanded $M=3$ response can be written
\begin{equation}
 G_3(k)=
 \frac{\bar\lambda_{+2}\bar\lambda_{+3}+(v_F^2/4)k^2}
 {\bar\lambda_{+1}\bar\lambda_{+2}\bar\lambda_{+3}
 +(v_F^2/4)k^2(\bar\lambda_{+1}+\bar\lambda_{+3})}.
 \label{eq:G3finiteB}
\end{equation}
In the collisionless high-field limit, its denominator vanishes at $kR_c=\sqrt6$ and its numerator at $kR_c=\sqrt{24}$.  We call a zero of the current response an \emph{antiresonance}.  The first value is a current pole of the $M=3$ rational approximant; the second is its first antiresonance.  Neither value is the exact kinetic location, and the fourth-order expansion cannot control wave numbers of this size.

\subsection{Exact collisionless hierarchy}

The contrast with the zero-field result in Eq.~\eqref{eq:B0exactTail} is instructive.  At $B=0$, an equal-rate tail is self-similar in harmonic index and closes through one algebraic fixed point.  At finite field, the precession term $\ii m\omega_c$ makes the recurrence explicitly index dependent and reorganizes the finite-$k$ response into Bessel functions.

Set
\begin{equation}
 x=kR_c=\frac{k v_F}{\omega_c}.
 \label{eq:xBessel}
\end{equation}
In the collisionless static limit, the positive-harmonic tail obeys
\begin{equation}
 m f_m+\frac{x}{2}(f_{m-1}+f_{m+1})=0,
 \qquad m\ge1.
 \label{eq:BesselRecurrence}
\end{equation}
The regular solution is
\begin{equation}
 f_m\propto(-1)^mJ_m(x),
 \label{eq:BesselSolution}
\end{equation}
by the standard Bessel recurrence.  The exact tail denominator is
\begin{equation}
 d_m=\ii\frac{v_Fk}{2}\frac{J_{m-1}(x)}{J_m(x)}.
 \label{eq:tailDenominator}
\end{equation}
For a source in $m=1$,
\begin{equation}
 \boxed{
 G_\infty(k)
 =\frac{2J_1(x)}{\ii\omega_c xJ_0(x)}.
 }
 \label{eq:exactBessel}
\end{equation}
The physical current poles are the zeros of $J_0(x)$, while the zeros of $J_1(x)$ are antiresonances where the current response vanishes.  Cyclotron-harmonic Bessel structures are classical in magnetized-plasma response and underlie Bernstein modes~\cite{Bernstein1958}.  The object here is different: the external frequency is fixed at $\Omega=0$, and the response is resolved in the spatial variable $kR_c$.  The closure-specific statement is that finite angular truncations approximate this spatial pole--zero structure.

If every harmonic in the tail has a common collision rate $\gamma$, the recurrence has complex order~\cite{Watson1944}.  With $\delta=\gamma/\omega_c$,
\begin{equation}
 G_\infty(k)
 =\frac{J_{1-\ii\delta}(x)}
 {\ii(v_Fk/2)J_{-\ii\delta}(x)}.
 \label{eq:complexOrderBessel}
\end{equation}
Collisions move the real-axis poles and zeros into broadened complex features.  More general angular rate spectra retain the continued-fraction structure but no longer collapse to one Bessel function of complex order.

\subsection{Finite closures as rational approximants}

\begin{figure*}[t]
 \centering
 \includegraphics[width=0.94\textwidth]{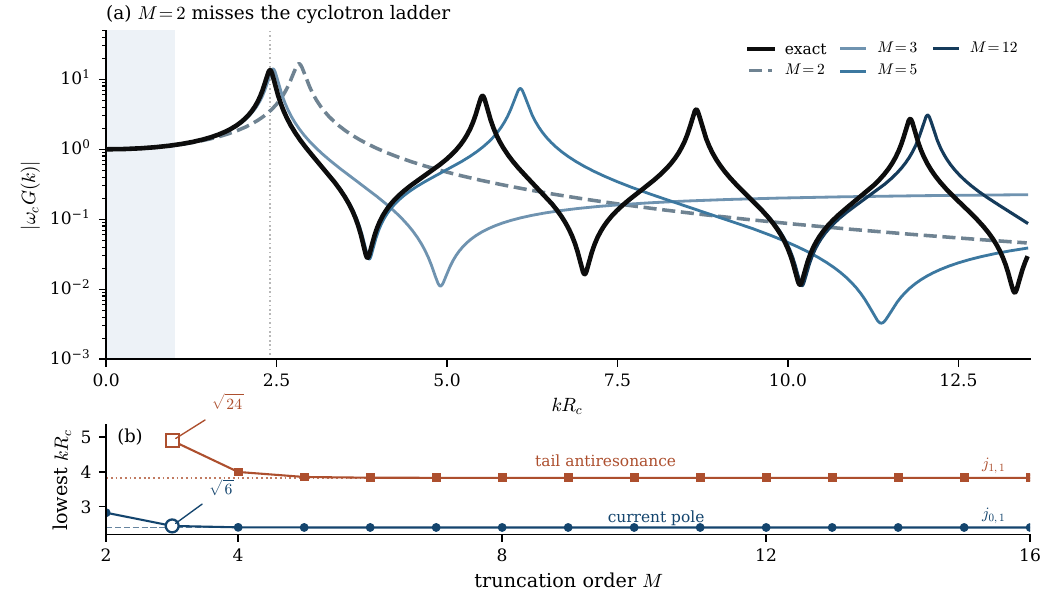}
 \caption{\label{fig:bessel}\textbf{Finite closures as rational approximants to the high-field kinetic response.}  \textbf{(a)} Magnitude of the dc helicity-resolved response for equal-rate broadening $\gamma/\omega_c=0.04$.  The black curve is the full complex-order Bessel result; the dashed $M=2$ curve and solid $M=3,5,12$ curves are finite closures.  The shaded interval $kR_c<1$ has $|\Xi_+|<0.042$, and the vertical guide marks $j_{0,1}$.  \textbf{(b)} Lowest current-pole and response-zero (antiresonance) branches.  The guides are $j_{0,1}$ and $j_{1,1}$; the open $M=3$ markers are $\sqrt6$ and $\sqrt{24}$.}
\end{figure*}

Finite closures are convergents of the continued fraction associated with the Bessel ratio in Eq.~\eqref{eq:exactBessel}.  Their approximation structure can be made explicit.  For a collisionless closure at order $M$, write the tail denominator as
\begin{equation}
 d_m^{(M)}=\ii\omega_c\frac{D_m^{(M)}}{D_{m+1}^{(M)}},
 \qquad
 D_{M+1}^{(M)}=1,
 \quad D_M^{(M)}=M.
 \label{eq:continuantDefinition}
\end{equation}
The continuants obey
\begin{equation}
 D_m^{(M)}=mD_{m+1}^{(M)}-\frac{x^2}{4}D_{m+2}^{(M)}.
 \label{eq:continuantRecurrence}
\end{equation}
After the rescaling
\begin{equation}
 \widehat D_m^{(M)}=\left(\frac{2}{x}\right)^{M+1-m}D_m^{(M)},
\end{equation}
one obtains
\begin{equation}
 \boxed{
 \widehat D_m^{(M)}
 =\frac{2m}{x}\widehat D_{m+1}^{(M)}
 -\widehat D_{m+2}^{(M)}.
 }
 \label{eq:LommelRecurrence}
\end{equation}
Up to normalization and reversal of the finite index, Eq.~\eqref{eq:LommelRecurrence} is the Lommel-polynomial recurrence associated with the same three-term chain as the Bessel functions~\cite{Watson1944}.  The finite current responses are therefore ratios of the corresponding Lommel continuants.  Classical results imply branchwise convergence of their zeros to the associated Bessel zeros~\cite{FeinsilverSchott1992}; Fig.~\ref{fig:bessel} illustrates this theorem rather than serving as its sole evidence.  Increasing $M$ introduces further pole--zero pairs approximating higher Bessel branches.  A selector based only on the globally largest feature can consequently jump between branches even though each fixed low branch converges.  The numerical root table is given in Supplemental Material, Sec.~\ref{sec:suppBessel}.

The conclusion is deliberately limited.  The fourth-order symbol in Eq.~\eqref{eq:principalSymbol}, the circular factorization in Eq.~\eqref{eq:bilaplacianIdentity}, and the chiral coefficients in Eq.~\eqref{eq:helicityCoefficients} are controlled low-$k$ statements satisfying Eqs.~\eqref{eq:XiControl} or \eqref{eq:XiHelicity}.  At $kR_c=\order(1)$, the kinetic response is the Bessel or continued-fraction hierarchy.  A finite-order resonance inferred from one denominator is a property of the rational approximant unless its branch convergence is demonstrated.

\section{Discussion}
\label{sec:discussion}

\subsection{What is classical and what is specific}

The bare existence of a fourth-order kinetic correction is not the result.  Burnett and super-Burnett terms are classical, and their unregularized finite-order equations can be unstable~\cite{Burnett1936,Bobylev1982,Uribe2000}.  Regularized moment systems such as R13 cure related pathologies through a structured closure that remains linearly stable beyond the regime of the bare Burnett polynomial~\cite{StruchtrupTorrilhon2003}.  The construction here is analogous in purpose but not in derivation: the rational expression is the exact Schur complement of a specified angular sector, not a phenomenological regularization.  What is specific here is the operator path, coefficient, symmetry factorization, and finite-$k$ completion.

First, the coefficient is generated by the leading Schur-complement feedback through the lowest discarded harmonic of a two-dimensional angular hierarchy,
\begin{equation}
 \kappa_4=\frac{\nu^2}{\gamma_3}.
 \label{eq:discussionKappa}
\end{equation}
It is controlled by the relaxation of a definite omitted Fermi-surface harmonic.  Second, circular symmetry factorizes the entire path exactly into $\Delta_{J-1}^2$ within the same $J$ block.  Third, the closed-form Green function of the unexpanded $M=3$ Schur complement proves a direction of spectral transfer rather than merely displaying a higher derivative.  Fourth, a magnetic field turns the coefficient into a chiral object with a closed-form sign reversal and an odd-mode-enhanced scaling function.  Finally, the finite-$k$ analytic structure itself is field dependent.  At zero field, positive rates exclude real-wave-number poles and antiresonances for any convergent hierarchy; the equal-rate benchmark has a square-root branch cut.  The high-field hierarchy instead supplies a Bessel pole--zero ladder.

These statements are compatible with systematic and regularized moment theories~\cite{Levermore1996,StruchtrupTorrilhon2003} but answer a narrower question.  This is a semiclassical angular-hierarchy calculation, not a universal theory of moments.  Its value is that one discarded mode can be followed all the way from a microscopic relaxation rate to a coordinate-independent operator, a circular Green function, and a chiral finite-field response.  It identifies the first term that distinguishes the source-to-current response after lower source moments have been reparametrized, and it shows how that obstruction is represented by circular symmetry and magnetic helicity.

\subsection{Relation to tomographic-transport studies}

This closure statement is distinct in kind from recent studies of the same odd-harmonic physics.  High-order cyclotron-resonance analyses test the tomographic Fermi-liquid hypothesis through the harmonic structure of electromagnetic absorption and magnetoconductivity~\cite{Moiseenko2025}, while finite-field studies characterize how a magnetic field suppresses tomographic transport through the static conductivity~\cite{Rostami2025}.  Those works evaluate spectroscopic or transport observables within a specified kinetic model.  The present result is complementary: it identifies the operator omitted by a Navier--Stokes-level moment closure, fixes its coefficient $\kappa_4=\nu^2/\gamma_3$ from the relaxation rate of the lowest discarded harmonic, and gives its exact circular and chiral representations.

The Bessel ladder of Sec.~\ref{sec:bessel} is not introduced as a new resonance prediction; it is the exact finite-$k$ completion showing that finite moment closures are rational approximants to the collisionless high-field response.  Likewise, $\kappa_\pm$ is not proposed as an additional magnetotransport observable but is the closure-error coefficient itself.  The connection to the tomographic literature is therefore physical rather than competitive: the same anomalously long-lived $m=3$ harmonic that underlies tomographic transport parametrically enhances the obstruction identified here through the $1/\gamma_3$ scaling in Eq.~\eqref{eq:kappaMaximum}.

\subsection{Connection to circular magnetotransport}

The circular density-perturbation problem studied in Ref.~\cite{ParasharFogler2026} is organized by the same conserved total angular momentum.  A uniform drive occupies $J=\pm1$, while the radial structure controls the crossover among diffusive, viscous, and spiraling response, as well as the long-distance Landauer dipole.  The present result adds a closure statement to that geometry: eliminating $m=3$ does not create a new angular sector; it modifies the radial spectrum within $J=\pm1$ through the unexpanded operator in Eq.~\eqref{eq:exactCircularSchur}.

At zero field, the correction preferentially reweights higher radial multipoles.  At finite field, its chiral component can oppose or reinforce Hall viscosity depending on $\omega_c$, and a long-lived $m=3$ mode moves the loss of second-order closure control to the scale $\ellthree=v_F/\gamma_3$.  These are operator statements about the diffusive--hydrodynamic--kinetic crossover, not claims tied to a particular probe or inversion protocol.

Two coefficient-level signatures follow without specifying an apparatus.  The $k^4$ Hall correction reverses sign at the field $\omega_{c,0}$ in Eq.~\eqref{eq:signReversal}, while odd-mode protection enhances its magnitude and shifts its competition with the $k^2$ Hall-viscous term to wave numbers $k\ellthree=\order(1)$.  Any response that resolves the wave-number dependence of the transverse current can, in principle, distinguish these features by their fourth-order scaling and field-tunable sign, without interpreting $\kappa_\pm$ as a new independent hydrodynamic coefficient.

\subsection{Scope and extensions}

Four restrictions should remain explicit.  First, all coefficients, Green functions, and resummations presented here are dc results with $\Omega=0$.  Retaining $-\ii\Omega$ in $\lambda_m$ produces frequency-dependent memory and spectroscopic poles, but that extension addresses a different question from the static closure obstruction studied here.

Second, the exact circular factorization uses constant isotropic kinetic coefficients.  If $\gamma_m$ or $v_F$ varies with radius, the fourth-order principal symbol is unchanged locally, but commutators with the coefficients generate additional lower-order radial terms.  Boundaries similarly affect the radial spectrum and Green function but not the local principal symbol.

Third, rotational symmetry is used to conserve $J$.  For a noncircular perturbation, different $J$ blocks mix.  The invariant symbol $-\kappa_4|\bm\xi|^4$ survives, while geometry enters through lower-order couplings and boundary conditions.  This makes anisotropic perturbations and many-defect superposition natural applications rather than separate closure problems.

Fourth, the collision operator has been represented by one angular rate $\gamma_m$ per harmonic.  This is sufficient for the structural Schur-complement result.  A microscopic collision operator with radial energy eigenmodes replaces $\lambda_m$ by a block in that radial index.  The same projection algebra then produces matrix-valued viscosities and fourth-order coefficients.  The scalar enhancement in Eq.~\eqref{eq:kappaMaximum} applies when the driven deformation has dominant overlap with an isolated slow $m=3$ eigenmode, as occurs in the low radial mode identified in microscopic two-dimensional Fermi-liquid calculations~\cite{NilssonGranHofmann2025}.

\section{Conclusion}
\label{sec:conclusion}

The lowest discarded angular harmonic leaves a coordinate-independent trace in the low-moment transport operator.  In Fourier variables that trace is the fourth-order principal symbol $-\kappa_4|\bm\xi|^4$.  In circular coordinates it factorizes exactly into $-\kappa_4\Delta_{J-1}^2$ inside the same conserved total-angular-momentum block.  The radial Green function of the unexpanded $M=3$ Schur complement is a local source copy plus a screened Bessel kernel, and its gain relative to the $M=2$ closure increases monotonically with radial wave number.  A magnetic field complexifies the coefficient, produces a field-driven Hall sign reversal, and permits a $1/\gamma_3$ enhancement when the lowest discarded odd harmonic is long lived.

The finite-wave-number completion is equally important.  The fourth-order coefficient is a Taylor coefficient of an exact Schur complement, controlled by $\Xi\ll1$, rather than a globally valid finite-wave-number equation.  At zero field, positivity of the continued fraction excludes a real-wave-number pole--antiresonance ladder, while the equal-rate benchmark resums to a square-root branch cut.  In the collisionless high-field limit, the complete hierarchy instead resums to a Bessel ratio, and finite closures approximate its pole--antiresonance ladder.  This separation between controlled principal symbol and field-dependent finite-$k$ resummation gives a stable formulation of closure failure in circular kinetic magnetotransport.

\section*{Data availability}
No datasets were generated or analyzed in this study.  All figures are
computed directly from the closed-form expressions given in the text, and
the numerical root values underlying Fig.~\ref{fig:bessel}(b) are tabulated
in the Supplemental Material.

\begin{acknowledgments}
The author thanks M. M. Fogler for prior collaboration on inhomogeneous magnetotransport and for discussions that motivated the circular formulation. Questions posed by Daniel Arovas during the author's thesis defense helped motivate the problem studied here.
\end{acknowledgments}

\bibliographystyle{apsrev4-2}
\bibliography{references}

\clearpage
\onecolumngrid
\setcounter{section}{0}
\setcounter{subsection}{0}
\setcounter{equation}{0}
\setcounter{figure}{0}
\setcounter{table}{0}
\renewcommand{\thesection}{S\arabic{section}}
\renewcommand{\thesubsection}{S\arabic{section}.\Alph{subsection}}
\renewcommand{\theequation}{S\arabic{equation}}
\renewcommand{\thefigure}{S\arabic{figure}}
\renewcommand{\thetable}{S\arabic{table}}
\renewcommand{\theHsection}{S\arabic{section}}
\renewcommand{\theHsubsection}{S\arabic{section}.\Alph{subsection}}
\renewcommand{\theHequation}{S\arabic{equation}}
\renewcommand{\theHfigure}{S\arabic{figure}}
\renewcommand{\theHtable}{S\arabic{table}}

\begin{center}
{\large\bfseries Supplemental Material for\\[2pt]
``Fourth-order closure obstruction and chiral nonlocality in circular kinetic magnetotransport''}\\[8pt]
P. Shubham Parashar\\
\textit{Department of Physics, University of California, San Diego, La Jolla, California 92093, USA}\\[4pt]
(Dated: July 2026)
\end{center}

This Supplemental Material records the derivations and consistency checks supporting the main text: the Schur complement and low-$q$ source-to-current expansion, zero-field continued fractions, circular factorization, finite-field helicity coefficients, the radial Green kernel, and the high-field Bessel resummation.  The final section contains the declared smoothed-ring illustration, its mode-weight redistribution, and the corresponding numerical checks.

\twocolumngrid
\section{Schur complement and the shortest omitted path}
\label{sec:suppSchur}

Let $P$ project onto the retained angular harmonics and $Q=1-P$.  Decompose the linear kinetic operator as
\begin{equation}
 \mcL=
 \begin{pmatrix}
  \mcL_{PP} & \mcL_{PQ}\\
  \mcL_{QP} & \mcL_{QQ}
 \end{pmatrix},
 \qquad
 \begin{pmatrix}\phi_P\\\phi_Q\end{pmatrix},
\end{equation}
with source restricted to the retained sector.  Solving the $Q$ equation gives
\begin{equation}
 \phi_Q=-\mcL_{QQ}^{-1}\mcL_{QP}\phi_P,
\end{equation}
and hence
\begin{equation}
 \boxed{
 \mcL_{\rm eff}
 =\mcL_{PP}-\mcL_{PQ}\mcL_{QQ}^{-1}\mcL_{QP}.
 }
 \label{eq:suppSchur}
\end{equation}
For an $M=2$ closure, $P=P_2$ retains $|m|\le2$.  Streaming is tridiagonal in angular momentum, so the shortest path leaving and returning to the current sector is
\begin{equation}
 1\rightarrow2\rightarrow3\rightarrow2\rightarrow1.
 \label{eq:suppPath}
\end{equation}
The four streaming vertices in Eq.~\eqref{eq:suppPath} produce a fourth-order current-sector symbol.  This is a path-counting statement and does not depend on a coordinate representation.

For a plane wave, define
\begin{equation}
 t_q=\frac{v_Fq}{2},
 \qquad
 \lambda_m=\gamma_m-\ii\Omega+\ii m\omega_c.
\end{equation}
The positive-harmonic tail satisfies
\begin{equation}
 \lambda_m f_m+\ii t_q(f_{m-1}+f_{m+1})=0.
\end{equation}
Eliminating $f_2$ in the $M=2$ hierarchy yields
\begin{equation}
 \Lambda_2(q)=\lambda_1+\frac{t_q^2}{\lambda_2}.
\end{equation}
Eliminating $f_3$ first and then $f_2$ gives
\begin{equation}
 \boxed{
 \Lambda_3(q)
 =\lambda_1+\frac{t_q^2}{\lambda_2+t_q^2/\lambda_3}.
 }
 \label{eq:suppLambda3}
\end{equation}
The positive sign in the denominator follows from the two factors of $\ii$ carried by streaming.  At $B=\Omega=0$,
\begin{align}
 \Lambda_3(q)
 &=\gamma_1+\frac{(v_F^2/4)q^2}{\gamma_2+(v_F^2/4\gamma_3)q^2}\nonumber\\
 &=\gamma_1+\nu q^2-\kappa_4q^4+\order(q^6),
\end{align}
with
\begin{equation}
 \boxed{
 \nu=\frac{v_F^2}{4\gamma_2},
 \qquad
 \kappa_4=\frac{v_F^4}{16\gamma_2^2\gamma_3}
 =\frac{\nu^2}{\gamma_3}.
 }
\end{equation}
The minus sign of the fourth-order term is therefore the low-wave-number trace of the anti-Hermitian streaming chain, not a sign convention imposed at the constitutive level.

\section{Zero-field equal-rate benchmark}
\label{sec:suppB0Tail}

Assume $B=\Omega=0$ and $\gamma_m=\gamma$ for every unresolved harmonic $m\ge2$.  If $X_m$ denotes the self-energy of the tail beginning at harmonic $m$, then
\begin{equation}
 X_m=\frac{t_q^2}{\gamma+X_{m+1}}.
 \label{eq:suppTailRecursion}
\end{equation}
The equal-rate semi-infinite tail is invariant under $m\mapsto m+1$, so its fixed points satisfy
\begin{equation}
 X_\infty=\frac{t_q^2}{\gamma+X_\infty}.
 \label{eq:suppTailFixedPoint}
\end{equation}
Every finite convergent is nonnegative for real $q$ and vanishes with $t_q^2$.  Equivalently, Pincherle's theorem selects the minimal solution of the associated three-term recurrence, namely the branch continuously connected to those convergents~\cite{Gautschi1967}.  The second algebraic root tends to $-\gamma$ as $q\to0$ and cannot represent a tail self-energy.  The selected solution is therefore
\begin{equation}
 \boxed{
 X_\infty(q)=\frac{-\gamma+\sqrt{\gamma^2+4t_q^2}}{2}
 =\frac{\sqrt{\gamma^2+v_F^2q^2}-\gamma}{2}.
 }
 \label{eq:suppTailSolution}
\end{equation}
Consequently,
\begin{align}
 \Lambda_\infty(q)
 &=\gamma_1+X_\infty(q)\nonumber\\
 &=\gamma_1+\frac{v_F^2}{4\gamma}q^2
 -\frac{v_F^4}{16\gamma^3}q^4+\order(q^6).
 \label{eq:suppTailExpansion}
\end{align}
Thus $\nu=v_F^2/(4\gamma)$ and $\kappa_4=\nu^2/\gamma$, in agreement with the fourth-order Schur-complement coefficient.  The square root has branch points at $q=\pm\ii\gamma/v_F$.  This equal-rate benchmark therefore has a branch-cut completion.  It is not representative of odd-mode protection because $\gamma_2=\gamma_3$; for unequal rates the off-axis singularity structure can be more complicated.  The general zero-field result is instead the real-axis positivity bound in Eq.~\eqref{eq:B0positivityBounds}.

\section{Low-$q$ source-to-current expansion and source-shape directions}
\label{sec:suppSource}

Let the inversion-symmetric source have a regular expansion
\begin{equation}
 S(q)=S_0(1-s_2q^2+s_4q^4+\cdots).
\end{equation}
The inverse current operator is
\begin{align}
 \frac{1}{\Lambda_3(q)}
 &=\frac1{\gamma_1}
 \frac1{1+(\nu/\gamma_1)q^2-(\kappa_4/\gamma_1)q^4+\cdots}\nonumber\\
 &=\frac1{\gamma_1}
 \left[
 1-\alpha q^2+
 \left(\alpha^2+\frac{\kappa_4}{\gamma_1}\right)q^4
 +\order(q^6)
 \right],
 \label{eq:suppInverseJet}
\end{align}
where $\alpha=\nu/\gamma_1$.  Multiplication by the source gives
\begin{equation}
 \frac{j(q)}{S_0/\gamma_1}
 =1-(s_2+\alpha)q^2
 +\left[s_4+\alpha s_2+\alpha^2+\frac{\kappa_4}{\gamma_1}\right]q^4
 +\order(q^6).
 \label{eq:suppResponseJet}
\end{equation}
Thus the second-order response identifies only $s_2+\nu/\gamma_1$.  This degeneracy is independent of the coordinates used to represent the source.  The fourth-order operator coefficient becomes distinguishable after the allowed source-shape variations are fixed or projected out.  If $s_4$ is also unconstrained, it supplies an additional fourth-order source-shape direction; a calibrated source family fixes the relation between $s_2$ and $s_4$.

For a one-scale Gaussian source, $S(qR)=\exp[-(qR)^2/2]$, the tangent identity is exact over the finite wave-number band.  The overall source amplitude is immaterial and is set to unity.  With
\begin{equation}
 \mathcal B_2(q)=\frac{\exp[-(qR)^2/2]}{\gamma_1+\nu q^2},
 \qquad
 \alpha=\frac{\nu}{\gamma_1},
\end{equation}
one finds
\begin{equation}
 \boxed{
 \partial_{\ln\nu}\mathcal B_2
 =\frac{\alpha}{R^2}\partial_{\ln R}\mathcal B_2
 +\alpha^2\mathcal B_2\frac{q^4}{1+\alpha q^2}.
 }
 \label{eq:suppExactTangent}
\end{equation}
The first term is the exact source-width direction, while the second is the fourth-order tail independent of source width.  Equation~\eqref{eq:suppExactTangent} is one explicit realization of the separation in Eq.~\eqref{eq:suppResponseJet}.

\section{Principal symbol and variable coefficients}
\label{sec:suppSymbol}

For constant isotropic rates, the leading omitted current operator at $B=0$ is
\begin{equation}
 \delta\mcL_j^{(4)}=-\kappa_4\Delta^2.
\end{equation}
The principal symbol is obtained by replacing $\partial_a\mapsto\ii\xi_a$,
\begin{equation}
 \sigma_4(\delta\mcL_j)(\bm x,\bm\xi)
 =-\kappa_4(x)(g^{ab}\xi_a\xi_b)^2.
 \label{eq:suppPrincipal}
\end{equation}
In flat space this reduces to $-\kappa_4|\bm\xi|^4$.  Under a smooth coordinate change, Eq.~\eqref{eq:suppPrincipal} transforms as a scalar principal symbol.  If $\gamma_m$, $v_F$, or therefore $\kappa_4$ vary in space, commuting derivatives through the coefficients generates terms with at most three derivatives on the distribution.  Such terms modify the subprincipal structure but not Eq.~\eqref{eq:suppPrincipal}.  The same separation holds for boundary terms: they determine the radial spectrum and domain of the operator, not its local fourth-order principal symbol.

\section{Circular streaming ladders and exact factorization}
\label{sec:suppCircular}

For one total-angular-momentum block,
\begin{equation}
 f_m^{(J)}(r)e^{\ii(J-m)\varphi}e^{\ii m\vartheta},
\end{equation}
the two polar pieces of $\hat{\bm p}\cdot\nabla$ shift $m\to m\pm1$ while preserving $J$.  Define
\begin{equation}
 \mcD_m^+=\partial_r+\frac{J-m}{r},
 \qquad
 \mcD_m^-=\partial_r-\frac{J-m}{r}.
\end{equation}
Let $n=J-m$.  Acting on a test function $u(r)$,
\begin{align}
 \mcD_{m+1}^-\mcD_m^+u
 &=\left(\partial_r-\frac{n-1}{r}\right)
   \left(\partial_r+\frac{n}{r}\right)u\nonumber\\
 &=u''+\frac1r u'-\frac{n^2}{r^2}u
 =\Delta_nu,
\end{align}
where
\begin{equation}
 \Delta_n=\partial_r^2+\frac1r\partial_r-\frac{n^2}{r^2}.
\end{equation}
Thus
\begin{equation}
 \boxed{
 \mcD_{m+1}^-\mcD_m^+=\Delta_{J-m}.
 }
 \label{eq:suppFirstLadder}
\end{equation}
The load-bearing fourth-order identity is
\begin{equation}
 \boxed{
 \mcD_2^-\Delta_{J-2}\mcD_1^+=\Delta_{J-1}^2.
 }
 \label{eq:suppBilaplacian}
\end{equation}
A direct verification follows by setting $n=J-1$ and applying both sides to $u(r)$.  The left-hand side becomes
\begin{align}
 &\left(\partial_r-\frac{n-1}{r}\right)
 \left(\partial_r^2+\frac1r\partial_r
 -\frac{(n-1)^2}{r^2}\right)\nonumber\\
 &\qquad\times
 \left(\partial_r+\frac{n}{r}\right)u\nonumber\\
 &=u^{(4)}+\frac{2}{r}u^{(3)}
 -\frac{1+2n^2}{r^2}u''\nonumber\\
 &\qquad+
 \frac{1+2n^2}{r^3}u'
 +\frac{n^2(n^2-4)}{r^4}u.
\end{align}
which is exactly $\Delta_n^2u$.

At $B=0$, the $m=3$ equation gives
\begin{equation}
 f_3^{(J)}
 =-\frac{v_F}{2\gamma_3}\mcD_2^+f_2^{(J)}.
\end{equation}
Substitution into the stress equation produces
\begin{equation}
 \left(\gamma_2-\frac{v_F^2}{4\gamma_3}\Delta_{J-2}\right)f_2^{(J)}
 =-\frac{v_F}{2}\mcD_1^+f_1^{(J)}.
\end{equation}
Eliminating $f_2^{(J)}$ yields the unexpanded $M=3$ current operator
\begin{equation}
 \boxed{
 \mcL_{j,3}^{(J)}
 =\gamma_1-\frac{v_F^2}{4}\mcD_2^-
 \left(\gamma_2-\frac{v_F^2}{4\gamma_3}\Delta_{J-2}\right)^{-1}
 \mcD_1^+.
 }
 \label{eq:suppCircularSchur}
\end{equation}
Expanding the inverse and using Eqs.~\eqref{eq:suppFirstLadder} and \eqref{eq:suppBilaplacian} gives
\begin{equation}
 \mcL_j^{(J)}
 =\gamma_1-\nu\Delta_{J-1}-\kappa_4\Delta_{J-1}^2+\order(\nabla^6).
\end{equation}
The correction remains in the same $J$ block.  The circular representation converts powers of wave number into powers of the radial Laplacian without introducing a new total-angular-momentum channel.

\section{Radial Green kernel and monotone spectral transfer}
\label{sec:suppGreen}

Set $z=k^2$.  The $M=2$ and unexpanded $M=3$ resolvents are
\begin{equation}
 R_2(z)=\frac1{\gamma_1+\nu z},
 \qquad
 R_3(z)=\frac1{\gamma_1+\nu z/(1+\nu z/\gamma_3)}.
\end{equation}
Writing the second as a rational function and matching its constant and pole parts gives
\begin{equation}
 \boxed{
 R_3(z)=\frac1{\gamma_1+\gamma_3}
 +\frac{\gamma_3^2}{\nu(\gamma_1+\gamma_3)^2}
 \frac1{z+\mu_3^2},
 }
\end{equation}
with
\begin{equation}
 \mu_3^2=\frac{\gamma_1\gamma_3}{\nu(\gamma_1+\gamma_3)}.
\end{equation}
The radial inverse of $k^2+\mu^2$ in angular channel $n$ is
\begin{equation}
 I_n(\mu r_<)K_n(\mu r_>)
\end{equation}
with measure $r'\dd r'$.  Hence
\begin{align}
 G_2^{(n)}(r,r')
 &=\frac1\nu I_n(\mu_2r_<)K_n(\mu_2r_>),\\
 G_3^{(n)}(r,r')
 &=\frac1{\gamma_1+\gamma_3}\frac{\delta(r-r')}{r}\nonumber\\
 &\quad+\frac{\gamma_3^2}{\nu(\gamma_1+\gamma_3)^2}
 I_n(\mu_3r_<)K_n(\mu_3r_>),
\end{align}
where $\mu_2^2=\gamma_1/\nu$.

The exact relative gain is
\begin{equation}
 Q(z)=\frac{R_3(z)}{R_2(z)}.
\end{equation}
Differentiation gives
\begin{equation}
 \boxed{
 Q'(z)=
 \frac{\nu^2z[2\gamma_1\gamma_3+\nu z(\gamma_1+\gamma_3)]}
 {[\gamma_1\gamma_3+\nu z(\gamma_1+\gamma_3)]^2}>0.
 }
 \label{eq:suppGainDerivative}
\end{equation}
For a discrete radial spectrum, let
\begin{equation}
 p_{2\ell}=\frac{|R_2(k_\ell^2)s_\ell|^2}
 {\sum_j|R_2(k_j^2)s_j|^2},
 \qquad
 p_{3\ell}=\frac{Q(k_\ell^2)^2p_{2\ell}}
 {\sum_jQ(k_j^2)^2p_{2j}}.
\end{equation}
For any increasing function $F(k^2)$,
\begin{align}
 \langle F\rangle_3-\langle F\rangle_2
 &=\frac{1}{2\langle Q^2\rangle_2}
 \sum_{\ell j}p_{2\ell}p_{2j}\nonumber\\
 &\quad\times
 [F(k_\ell^2)-F(k_j^2)]
 [Q(k_\ell^2)^2-Q(k_j^2)^2]\ge0.
 \label{eq:suppMonotoneProof}
\end{align}
Strict inequality holds when the source occupies at least two modes on which both $F$ and $Q$ differ.  This proves the transfer toward higher radial multipoles independently of the source used for illustration.

\section{Finite-field chiral coefficient}
\label{sec:suppChiral}

At dc,
\begin{equation}
 \nu_+=\frac{v_F^2}{4(a+2\ii\omega)},
 \qquad
 \kappa_+=\frac{v_F^4}{16(a+2\ii\omega)^2(c+3\ii\omega)},
\end{equation}
where $a=\gamma_2$, $c=\gamma_3$, and $\omega=\omega_c$.  With
\begin{equation}
 \nu_+=\ve-\ii\vH,
 \qquad
 \kappa_+=\ke-\ii\kH,
\end{equation}
one obtains
\begin{align}
 \ve&=\frac{v_F^2a}{4(a^2+4\omega^2)},\\
 \vH&=\frac{v_F^2\omega}{2(a^2+4\omega^2)},\\
 \ke&=\frac{v_F^4[c(a^2-4\omega^2)-12a\omega^2]}
 {16(a^2+4\omega^2)^2(c^2+9\omega^2)},\\
 \kH&=\frac{v_F^4\omega[3a^2+4ac-12\omega^2]}
 {16(a^2+4\omega^2)^2(c^2+9\omega^2)}.
\end{align}
The Hall coefficient changes sign at
\begin{equation}
 \omega_{c,0}^2=\frac{a(3a+4c)}{12}.
\end{equation}
The even coefficient changes sign at
\begin{equation}
 \omega_{c,{\rm e}}^2=\frac{ca^2}{4(c+3a)}.
\end{equation}
Both are coefficient-level zeros and remain finite for all real field when the relaxation rates are positive.

For the odd-mode asymptotics, set
\begin{equation}
 r=\frac{c}{a},
 \qquad
 y=\frac{\omega_c}{a}.
\end{equation}
The positive extrema of $\kH$ solve
\begin{align}
 0={}&1296u^3+(48r^2-720r-648)u^2\nonumber\\
 &-(48r^3+72r^2+36r+27)u+4r^3+3r^2,
 \label{eq:suppPeakPolynomial}
\end{align}
where $u=y^2$.  The root that tends to zero with $r$ has
\begin{equation}
 y_{\max}=\frac r3-\frac89r^3+\order(r^4),
\end{equation}
or
\begin{equation}
 \boxed{
 \omega_c^{\max}
 =\frac{\gamma_3}{3}
 -\frac{8\gamma_3^3}{9\gamma_2^2}
 +\order\!\left(\frac{\gamma_3^4}{\gamma_2^3}\right).
 }
\end{equation}
At this point,
\begin{equation}
 \boxed{
 \kH^{\max}
 =\frac{v_F^4}{32\gamma_2^2\gamma_3}
 \left[
 1+\frac{4\gamma_3}{3\gamma_2}
 +\order\!\left(\frac{\gamma_3^2}{\gamma_2^2}\right)
 \right].
 }
 \label{eq:suppPeakAmplitude}
\end{equation}
The leading divergence is $1/\gamma_3$.  The correction to the prefactor begins at first order in $\gamma_3/\gamma_2$.

Keeping $x=3\omega_c/\gamma_3$ fixed while $r\to0$ gives
\begin{equation}
 \frac{32\gamma_2^2\gamma_3}{v_F^4}\kH
 =\frac{2x}{1+x^2}+\order(r),
\end{equation}
which is the scaling collapse used in the main text.  At $x=1$,
\begin{equation}
 \vH=\frac{v_F^2\gamma_3}{6\gamma_2^2}[1+\order(r^2)],
\end{equation}
and the low-gradient competition is
\begin{equation}
 \frac{\kH k^4}{\vH k^2}
 =\frac3{16}(k\ell_3)^2[1+\order(r)],
 \qquad
 \ell_3=\frac{v_F}{\gamma_3}.
\end{equation}
At high field,
\begin{align}
 \nu_+&=-\frac{\ii v_F^2}{8\omega_c}
 +\frac{v_F^2\gamma_2}{16\omega_c^2}+\order(\omega_c^{-3}),\\
 \kappa_+&=\frac{\ii v_F^4}{192\omega_c^3}
 -\frac{v_F^4(3\gamma_2+\gamma_3)}{576\omega_c^4}
 +\order(\omega_c^{-5}).
\end{align}
Thus $|\kH k^4/(\vH k^2)|\sim(kR_c)^2/24$.  This ratio locates the eventual failure of the low-$k$ expansion; it does not license extrapolation of the fourth-order polynomial to $kR_c\simeq\sqrt{24}$.

\section{Bessel resummation and rational approximants}
\label{sec:suppBessel}

In the collisionless static high-field hierarchy, let
\begin{equation}
 x=kR_c=\frac{k v_F}{\omega_c}.
\end{equation}
The positive tail satisfies
\begin{equation}
 m f_m+\frac x2(f_{m-1}+f_{m+1})=0,
 \qquad m\ge1.
\end{equation}
The regular solution is $f_m=(-1)^mJ_m(x)$.  The exact tail denominator is
\begin{equation}
 d_m=\ii\frac{v_Fk}{2}\frac{J_{m-1}(x)}{J_m(x)},
\end{equation}
and a source in $m=1$ gives
\begin{equation}
 \boxed{
 G_\infty(k)=\frac{2J_1(x)}{\ii\omega_c xJ_0(x)}.
 }
\end{equation}
The current poles are the zeros of $J_0$, while the zeros of $J_1$ are antiresonances.  If every tail harmonic has one common rate $\gamma$, the Bessel order becomes complex,
\begin{equation}
 G_\infty(k)=
 \frac{J_{1-\ii\delta}(x)}
 {\ii(v_Fk/2)J_{-\ii\delta}(x)},
 \qquad
 \delta=\frac\gamma{\omega_c}.
\end{equation}

For collisionless truncation at $M$, define recursively
\begin{equation}
 d_M=\ii M\omega_c,
 \qquad
 d_m=\ii m\omega_c+\frac{(v_Fk/2)^2}{d_{m+1}}.
\end{equation}
The lowest branch-resolved roots are:
\begin{table}[t]
\caption{Lowest current pole and tail antiresonance of the collisionless continued fraction.  The limits are the first zeros $j_{0,1}=2.404825557\ldots$ and $j_{1,1}=3.831705970\ldots$.}
\label{tab:suppRoots}
\begin{ruledtabular}
\begin{tabular}{ccc}
$M$ & current pole & tail antiresonance\\
\hline
2  & 2.828427 & ---\\
3  & 2.449490 & 4.898979\\
4  & 2.408456 & 4.000000\\
5  & 2.405021 & 3.855979\\
6  & 2.404833 & 3.834290\\
7  & 2.404826 & 3.831904\\
9  & 2.404826 & 3.831706\\
12 & 2.404826 & 3.831706
\end{tabular}
\end{ruledtabular}
\end{table}
The $M=3$ values $\sqrt6$ and $\sqrt{24}$ are therefore a pole and an antiresonance of the lowest rational approximant.  They are not finite-wave-number predictions of the fourth-order coefficient.  Additional truncation orders add further pole--zero pairs; global-peak tracking may switch branches even while the low branches converge.

\section{Controlled smoothed-ring example}
\label{sec:suppNumerics}

\begin{figure*}[t]
 \centering
 \includegraphics[width=0.94\textwidth]{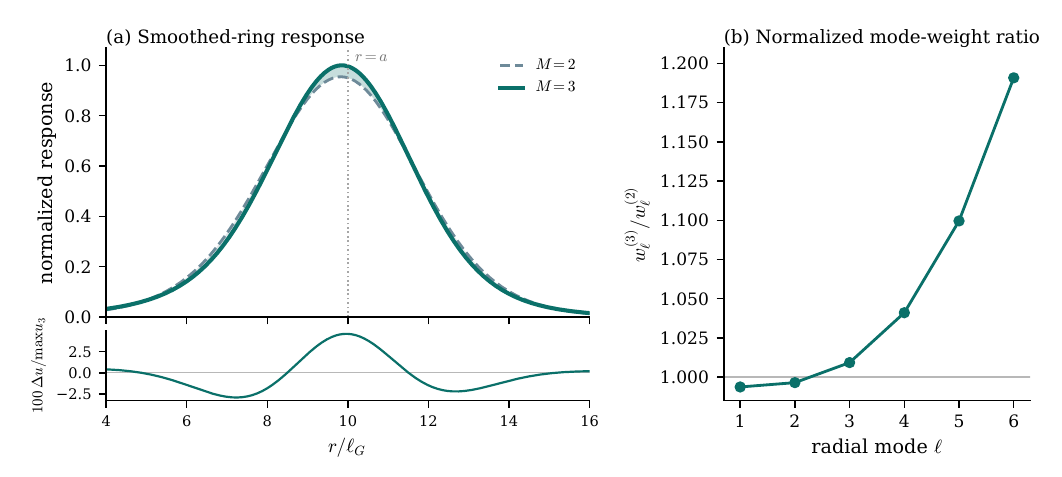}
 \caption{\label{fig:suppRing}\textbf{Smoothed-ring illustration of the radial reweighting.}  \textbf{(a)} Normalized $M=2$ and unexpanded $M=3$ responses for a Gaussian-smoothed ring with $a=10\ell_G$ and $\sigma=1.5\ell_G$; the lower strip shows their signed relative difference.  \textbf{(b)} Ratio of normalized mode weights for the six modes with $\Xi_\ell\le0.30$.  The example illustrates the monotone gain proved in the main text; the source-independent result is Eq.~\eqref{eq:gainDerivative}.}
\end{figure*}

The radial illustration uses
\begin{equation}
 \gamma_1=\gamma_2=1,
 \qquad
 \gamma_3=2,
 \qquad
 v_F=2,
\end{equation}
so that $\nu=1$ and $\kappa_4=1/2$.  The source is
\begin{equation}
 \widetilde s(k)=J_0(ka)e^{-\sigma^2k^2/2},
 \qquad
 a=10\ell_G,
 \qquad
 \sigma=1.5\ell_G,
\end{equation}
with $\ell_G=\sqrt{\nu/\gamma_1}$.  An auxiliary disk of radius $23.3296\ell_G$ discretizes the Hankel spectrum.  Six modes satisfy the declared control condition
\begin{equation}
 \Xi_\ell=\frac{\nu k_\ell^2}{\gamma_3}\le0.30.
\end{equation}
The real-space profile and normalized modal redistribution are shown in Fig.~\ref{fig:suppRing}; the numerical summaries are collected in Table~\ref{tab:suppRadial}.

\begin{table}[t]
\caption{Controlled radial-mode illustration for the smoothed-ring source.}
\label{tab:suppRadial}
\begin{ruledtabular}
\begin{tabular}{lc}
quantity & value\\
\hline
$\langle k^2\rangle_3/\langle k^2\rangle_2$ & 1.0439\\
$\langle k^4\rangle_3/\langle k^4\rangle_2$ & 1.0954\\
$M=2$ weight in modes $\ell\ge4$ & 0.0398\\
$M=3$ weight in modes $\ell\ge4$ & 0.0436\\
Hankel/real-space agreement, $M=2$ & $3.2\times10^{-5}$\\
Hankel/real-space agreement, $M=3$ & $1.4\times10^{-5}$
\end{tabular}
\end{ruledtabular}
\end{table}

\end{document}